\documentclass[12pt,a4paper]{article}
\usepackage{amsmath}
\usepackage{graphicx}
\usepackage{times}
\begin{document}
\begin{titlepage}
\title{Inelastic diffraction and role of reflective scattering at the LHC}
\author{ S.M. Troshin, N.E. Tyurin\\[1ex]
\small  \it Institute for High Energy Physics,\\
\small  \it Protvino, Moscow Region, 142281, Russia}
\normalsize
\date{}
\maketitle

\begin{abstract}
We  discuss the high-energy dependencies of  diffractive and non-diffractive inelastic cross-sections  in view of 
the recent LHC data   which  revealed a presence of the reflective scattering mode. 
\end{abstract}
\end{titlepage}
\setcounter{page}{2}
\section*{Introduction}
The  measurements performed by the experiments ALICE, ATLAS , CMS, LHCb and TOTEM  at the LHC have confirmed  
an increase with energy of the total, elastic and inelastic cross--sections, the trend  earlier observed  at lower energies 
(cf. \cite{lipari} for  interpretation of the new data  and the respective references). 
The results of these measurements are  bringing us  closer to a revelation of the elusive asymptotic regime of 
strong interactions.  

The analysis of the data on elastic scattering obtained by the TOTEM  at $\sqrt{s}=7$ TeV has revealed an existence  of the new regime in strong interaction dynamics,  related to transition to the new scattering mode described in  \cite{ph1,phl,degr,intja},
antishadowing or reflective scattering at very high energies. Experimentally, its appearance is manifested under a reconstruction of the elastic amplitude, elastic and inelastic
overlap functions in the impact parameter representation  \cite{alkin}. The nomenclature of reflective scattering is not a widely used one and  should be clarified in what follows.

The main issue of this note is related to discussion  of the reflective scattering mode, its influence and manifestation in the inelastic diffraction at the LHC.  
In particular,  an upper bound on the inelastic diffractive cross--section in the case when this mode starts
to be observed is obtained.
\section{Reflective and absorptive scattering modes}
The unitarity equation  in the impact parameter representation
assumes the two scattering modes, which can be designated as absorptive and reflective ones and the particular selection  will be described below. 
An attractive feature of the impact parameter picture  is diagonalisation of the unitarity equation  written for the elastic scattering amplitude $f(s,b)$, i.e.  
\begin{equation}\label{unit}
\mbox{Im} f(s,b)=|f(s,b)|^{2}+h_{inel}(s,b)
\end{equation}
at high energies with ${\cal O}(1/s)$ precision \cite{gold}, with $b$ being an impact parameter of the  colliding hadrons. 
The $|f(s,b)|^{2}$ is the elastic channel contribution, while the inelastic overlap function $h_{inel}(s,b)$ 
covers the sum of the contributions from  all the intermediate inelastic channels.
The elastic scattering $S$-matrix element is  related to the elastic scattering amplitude $f(s,b)$ by the equation $S(s,b)=1+2if(s,b)$ and
can be represented in the form
\[S(s,b)=\kappa(s,b)\exp[2i\delta(s,b)]\]
with the two real functions $\kappa(s,b)$ and $\delta(s,b)$. 
The function $\kappa$ ($0\leq \kappa \leq 1$) is called an absorption factor: its value $\kappa=0$ corresponds to a complete absorption of the initial state. 
At high   energies the real part of the scattering amplitude is small and can therefore be neglected,  i.e. this fact allows the substitution $f\to if$ in what
follows.  It also means that
the function $S(s,b)$ is real, but it does not have a definite sign, i.e. it can be positive or negative.

In fact, the choice of elastic scattering mode, namely, absorptive  or reflective one, depends on the sign of the function $S(s,b)$, i.e. on the phase $\delta(s,b)$
\cite{ttprd}. The standard assumption is that  $S(s,b)\to 0$  at the fixed impact parameter $b$ and $s\to \infty$. This  is called a black disk limit, and  the elastic scattering  is completely absorptive.  In this case the function $S(s,b)$ is
always non-negative. It also implies the limitation $f(s,b)  \leq 1/2$.  

There is  an  another option: the function $ S(s,b)\to -1$ at fixed $b$  and $s\to \infty$, i.e.  $\kappa \to 1$ and $\delta = \pi/2$. This phase can be interpreted as the geometric phase related to the presence of singularity \cite{intja,arh13}.

Thus, the function $S(s,b)$ can be negative in the certain region of $s$- and $b$ - values. It happens, in particular, in the Donnachie--Landshoff model (cf. \cite{dl} and the references therein) at the LHC energies. But, this model does not preserve unitarity, the value of  $|S(s,b)|$ eventually exceeds unity at fixed impact parameter when the collision energy being high enough,  violating that way a probability conservation. At the LHC energies the amplitude in this case, exceeds the black disk limit at small impact parameters, but, the amplitude still obeys the unitarity limitation  (cf. \cite{adm}). This is consistent with a principal conclusion of the model--independent treatment of the impact parameter dependencies performed in paper  \cite{alkin}.

The limiting case  $ S(s,b)\to -1$ at fixed $b$ can be interpreted as a pure reflective scattering using analogy with a reflection of the light wave in optics \cite{intja}. 
The appearance of the reflective scattering can be associated with increasing density of a scatterer with  energy. It can be said that this density goes beyond the critical value, corresponding to the black disk limit, and that  the scatterer starts to reflect the initial wave in addition to its absorption. 
The principal point of the reflective scattering mode is that  $1/2  < f(s,b) \leq 1$ and $0  > S(s,b) \geq -1$, as allowed by unitarity relation \cite{ph1,phl}. 
The selection of  absorptive or  reflective  scattering   leads to the different 
values for the ratio $ \sigma_{el}(s)/\sigma_{tot}(s)$ at the asymptotical energies, as it will be discussed in what follows.

Indeed, the arguments based on analyticity and unitarity of the scattering matrix have
lead to conclusion that the Froissart-Martin bound \cite{froi,martin} on the total cross-sections 
would be saturated at the asymptotic  energies \cite{kupsch}.
The functional energy dependence of the  total cross-sections is often taken to have a $\ln^2 s$-dependence at very
high energies, but 
the value of the factor in front of $\ln^2 s$ remains to be  an issue. 
The value of this factor is
related to the choice  of the upper limit for the partial amplitude (or the amplitude in the impact parameter representation). 
The value of this  limit may correspond to the
maximum of the inelastic channel contribution to the elastic unitarity, when
\begin{equation}\label{bd}
  \sigma_{el}(s)/\sigma_{tot}(s)\to 1/2,
\end{equation}
or it might correspond to a maximal value of the partial amplitude allowed by unitarity resulting in the asymptotical limit
\begin{equation}\label{rd}
  \sigma_{el}(s)/\sigma_{tot}(s)\to 1.
\end{equation}
The first option is to be an equivalent of a supposed absorptive nature of the scattering, while the second option
assumes an alternative which was interpreted as a reflective scattering (cf. \cite{intja} and the above discussion).
Assuming absorptive nature of  scattering the original Froissart-Martin bound on the total cross-sections 
has been improved and an upper bound on the total inelastic cross--section reduced by factor of 1/4 has been derived \cite{mart}.
For the modern status of the bound on the total cross--section and bound on the inelastic cross--section without unknown constants 
see the recent papers \cite{mart1} and \cite{mart2}.

It should be noted that the ratio $ \sigma_{el}(s)/\sigma_{tot}(s)$ is standing in front of $\ln^2 s$ in the asymptotical bound on the total 
cross-section \cite{roy}:
\begin{equation}\label{sing}
\sigma_{tot}(s)\leq \frac{4\pi}{t_0}\left(\frac {\sigma_{el}(s)}{\sigma_{tot}(s)}\right)\left[\ln\left(\frac{s}{\sigma_{el}(s)}\right)\right]^2
\left[1+\left(\frac{\mbox{Re} F(s, t=0)}{\mbox{Im} F(s, t=0)}\right)^2\right]^{-1}.
\end{equation}
We assumed for simplicity  that the scale of $s$ is to be determined by $s_0=1$ GeV$^2$, but in fact, this scale is an energy-dependent one and is determined by $ \sigma_{el}(s)$ as it is clear from Eq. (\ref{sing}), $\sqrt{t_0}$  is the mass of the lowest state in the $t$ channel\footnote{For most cases, 
${t_0}=4m^2_{\pi}$ .} and $F(s,t)$ is the elastic scattering amplitude related to $f(s,b)$ by the Fourier-Bessel transformation.

\section{Modified upper bound on the inelastic diffraction}
An assumption on absorptive nature of the scattering is a crucial issue for  the derivation of the Pumplin bound \cite{pumplin,miet}, the upper bound
for the cross-section of the inelastic diffraction:
\begin{equation}\label{pump}
 \sigma_{diff}(s,b)\leq \frac{1}{2}\sigma_{tot}(s,b)-\sigma_{el}(s,b),
\end{equation}
where  \[\sigma_{diff}(s,b)\equiv \frac{1}{4\pi}\frac{d\sigma_{diff}}{db^2}\] is the total cross--section of  all the inelastic diffractive processes in the impact parameter 
representation and, respectively, 
 \[ \sigma_{tot}(s,b)\equiv \frac{1}{4\pi}\frac{d\sigma_{tot}}{db^2}\,\, ,\,\,  \sigma_{el}(s,b)\equiv \frac{1}{4\pi}\frac{d\sigma_{el}}{db^2}.\]
The Eq. (\ref{pump}) was obtained in the framework of the formalism where the inelastic diffraction is considered to be  a result of the different absorption 
of the relevant states \cite{pomer,gwalk}.
The respective bound on the non-diffractive cross-section is the following :
\begin{equation}\label{pumpnd}
 \sigma_{ndiff}(s,b)\geq \frac{1}{2}\sigma_{tot}(s,b)
\end{equation}
since  $\sigma_{ndiff}=\sigma_{inel}-\sigma_{diff}$. 
These relations, valid for   each value of the impact parameter of the collision, can be integrated 
over  $b$:
\begin{equation}\label{pumpint}
 \sigma_{diff}(s)\leq \frac{1}{2}\sigma_{tot}(s)-\sigma_{el}(s)\,\, \mbox{and} \,\, \sigma_{ndiff}(s)\geq \frac{1}{2}\sigma_{tot}(s).
\end{equation}

The experimental status of the  Eq. (\ref{pumpint}) at the LHC energies has been discussed recently in  \cite{lipari} and \cite{lands}. 
It was noted  that conclusion on the large magnitude of the inelastic diffraction cross-section follows from comparison
of the inelastic cross-section measurements performed by ATLAS  \cite{atlasi} and CMS  \cite{cmsi}  with the TOTEM. In order to
reconcile the data of all experiments one needs to assume large value for $\sigma_{diff}(s)$ and essential contribution from the low--mass region.
As it was noted in \cite{lands}, an account for the contribution from this region would lead to a resolution of the inconsistency in the 
different experimental results noted in \cite{khoze}.

Thus, the data obtained at the LHC demonstrate an approximate
energy--independence of the ratio
$ \sigma_{diff}(s)/\sigma_{inel}(s)$   \cite{alice}.
At $ \sqrt{s}= 7$ TeV this ratio   is about  1/3.
The ratio $ \sigma_{diff}(s)/\sigma_{el}(s)$   is approximately equal to unity and 
\begin{equation}\label{bpump}
  [\sigma_{el}(s)+\sigma_{diff}(s)]/\sigma_{tot}(s)= 0.495^{+0.05}_{-0.06}.
  \end{equation}
The above numbers have been taken  from  \cite{lipari}.

But, in the framework of the absorptive scattering,  Eqs. (\ref{bd}) and (\ref{pumpint}) should be fulfilled
simultaneously if the black disk limit is taking place asymptotically, i.e. 
\begin{equation}\label{bdin}
  \sigma_{inel}(s)/\sigma_{tot}(s)\to 1/2
\end{equation}
while
\begin{equation}\label{bdind}
  \sigma_{diff}(s)/\sigma_{tot}(s)\to 0
\end{equation}
and
\begin{equation}\label{bdindi}
  \sigma_{diff}(s)/\sigma_{inel}(s)\to 0
\end{equation}
at $s \to \infty$.

The  limits Eqs. (\ref{bdin}-\ref{bdindi}) are in contradiction. 
Indeed,  $\sigma_{diff}(s)$ should be, by definition\footnote{A common approach associates dynamics of the inelastic diffraction processes with one or several  Pomeron exchanges. Cf.  \cite{lands,pred} for discussion.}, a main part of the inelastic
cross--section $\sigma_{inel}(s)$. 
In contrast to this definition and the available  data, one should conclude from Eq. (\ref{bdindi}) that the inelastic diffraction is, in fact,  a sub-leading mechanism 
 in the increase of the inelastic cross-section and the main role in this increase
 belongs to the non-diffractive inelastic processes. 
 The above statement is difficult to conform to the existing experimental  trends observed at the LHC.

There is no such an apparent contradiction  in the approach   assuming saturation of the unitarity limit as it was discussed above. Indeed,
the assumption that unitarity limit is to be saturated asymptotically  leads to a  slower increase of the inelastic cross-section, i.e.
at $s\to \infty$
\begin{equation}\label{bdind0}
  \sigma_{inel}(s)/\sigma_{tot}(s)\to 0.
\end{equation}
 It allows one to keep considering the inelastic diffraction as a leading mechanism  responsible for the inelastic cross--section growth.
In this approach the ratio  of the elastic to total cross-section Eq. (\ref{rd}) corresponds to  energy increase of the total inelastic cross-section slower 
than  $\ln^2 s$ while both Eqs. (\ref{rd}) and  (\ref{bdind0}) take place. And the available experimental data are consistent with decreasing
 ratio $ \sigma_{inel}(s)/\sigma_{tot}(s)$ when the energy increases.

The model-independent  reconstruction of the impact--parameter dependent quantities from the experimental data demonstrates that the black disk limit has been exceeded in the elastic scattering at small values of $b$ \cite{alkin}. In fact, the elastic scattering $S$-matrix 
element $S(s,b)\equiv 1-2f(s,b)$, where  the elastic amplitude $f(s,b)$  is considered to be a real function, is negative at $0<b<0.2$ fm and crosses zero
at $b=0.2$ fm at $\sqrt{s}=7$ TeV. In particular, this is consistent with the result  of the Tevatron data analysis \cite{girom}. 

The possibility of going beyond the black disk limit  was discussed in the  framework of the rational 
unitarization and the CDF data obtained at Tevatron in \cite{phl}. It should be noted that the value of $\mbox{Im} f(s,b=0)$ 
has increased from $0.36$  (CERN ISR) to $0.492\pm 0.008$ (Tevatron)  and it is close to exceeding or saturation  the black disk limit in this energy domain\cite{girom}. As it was noted \cite{ph1,phl}, exceeding the black disk limit turns the derivation of the Pumplin bound to loose its ground.
In fact, this bound is not valid in the range of the small and moderate values of the  impact parameter, where the absorptive approach ceases to be applicable. 

The Pumplin bound can easily be rewritten in terms of $S(s,b)$  in the form
\begin{equation}\label{pbsf}
 \sigma_{diff}(s,b)\leq \frac{1}{4}S(s,b)[1-S(s,b)]. 
 \end{equation}
This inequality clearly indicates that this relation cannot be applied in the region where $S(s,b)$ is negative. 
This region is determined by the interval $0<b<R(s)$, where $R(s)$ is the solution of the equation $S(s,b)=0$. In the above mentioned impact parameter
range the obvious restriction
\begin{equation}\label{pbsb}
 \sigma_{diff}(s,b)\leq \sigma_{inel}(s,b)
 \end{equation}
can only  be applied. In case of reflective scattering this obvious restriction is not a completely trivial in view that $\sigma_{inel}(s,b)$ 
has a peripheral impact parameter dependence.
But, at $b\geq R(s)$ the scattering is absorptive and, therefore, the original  bound on the inelastic diffractive cross--section should be valid. 

However, the integrated over all values of $b$ relation
should be modified. Namely, in this case it is to be written in the form
\begin{equation}\label{pumprefl}
\bar \sigma_{diff}(s)\leq \frac{1}{2}\bar \sigma_{tot}(s)-\bar \sigma_{el}(s),
 \end{equation}
where  $\bar \sigma_i(s)$ are the reduced cross-sections:
 \[
 \bar \sigma_i(s)\equiv \sigma_i(s) -8\pi\int_0^{R(s)}bdb\sigma_i(s,b),
 \]
 and $i\equiv$ $diff, tot, el$, respectively. Combining Eqs. ( \ref{pbsb}) and  ( \ref{pumprefl}), the following inequalities relevant for the LHC energies, 
 can easily be obtained:
 \begin{equation}\label{brfd}
 \sigma_{diff}(s) \leq \sigma_{inel}(s)- 2\pi\int_{R(s)}^\infty bdb[1-S(s,b)]
 \end{equation}
and
\begin{equation}\label{brfnd}
 \sigma_{ndiff}(s) \geq 2\pi\int_{R(s)}^\infty bdb[1-S(s,b)].
 \end{equation}
 The function $S(s,b)$ can be reconstructed from the experimental data
 on $d\sigma/dt$ in elastic $pp$-scattering \cite{alkin}. 
\section{The model consideration} 
 The unitary model for the  $S(s,b)$ can also be used  to estimate qualitatively the dependencies of the cross-sections $ \sigma_{diff}(s)$ and 
 $\sigma_{ndiff}(s)$. The reflective scattering is a characteristic picture of the model. 
 It is based on the rational form of the unitarization and represents the function $S(s,b)$ 
 in the form:
 \begin{equation}\label{um}
 S(s,b)=   \frac{1-U(s,b)}{1+U(s,b)},
  \end{equation}   
 The $U(s,b)$ is the generalized reaction matrix element, which is considered to be an
input dynamical quantity and it is taken to be a real function. The form (\ref{um}) is a one-to-one transform and is easily
invertible.  The various dynamical models can be used for the function $U(s,b)$. To get the qualitative estimates we use the simplified form of this function
which conforms to rising total cross-section and analytical properties over the transferred momentum, i.e.
  \begin{equation}\label{umf}
 U(s,b) = g(s) \exp({-\mu b}),
  \end{equation}  
  where $g(s)\sim  s^\lambda $ , $\lambda$  and $\mu$ are the constants. 
  Eq. (\ref{umf}) can also  be motivated by the model proposed by Heisenberg in 1952 \cite{heis}.
  
  Then the following asymptotical dependencies will take place\footnote{The explicit expressions for $R(s)$ and $\sigma_{inel}(s)$  are the following    
 \[R(s)=\frac{1}{\mu}\ln g(s)\,\,\mbox{and}\,\, \sigma_{inel}(s)=\frac{8\pi}{\mu^2}\ln(1+g(s)). \]}:
   \begin{equation}\label{asym}
 \sigma_{tot}(s) \sim \ln^2 s, \, \, \sigma_{el}(s) \sim \ln^2 s,  \, \, \sigma_{inel}(s) \sim \ln s  \, \,    \mbox{and} \, \, R(s) \sim \ln s .
   \end{equation}  
   From Eq.  ( \ref{brfd}) it follows that for the ratio ${\sigma_{diff}(s) }/{\sigma_{inel}(s)}$ the inequality takes place  
   \begin{equation}\label{brfd1}
\frac {\sigma_{diff}(s) }{\sigma_{inel}(s)}\leq 1- \frac{2\pi}{\sigma_{inel}(s)}\int_{R(s)}^\infty bdb[1-S(s,b)].
 \end{equation}
From Eqs.  ( \ref{brfnd}) and  ( \ref{asym}) it follows that    $ \sigma_{ndiff}(s) \sim \ln s  $ and  second term in Eq. (\ref{brfd1})
tends to $1/2$ at $s\to \infty$. In general, to exclude a subleading role  of 
$ \sigma_{diff}(s)$, the factor in front of  $\ln s$ in  $\sigma_{ndiff}(s)$ should be different
 from the corresponding factor in  $\sigma_{inel}(s)$ and the asymptotical dependence of the inelastic diffractive cross-section would be
  $ \sigma_{diff}(s) \sim \ln s  $. Thus, in this approach both parts of $ \sigma_{inel}(s)$ would have similar asymptotical energy dependencies, which are proportional  to   $ \ln s  $, while the ratio of the inelastic diffractive to elastic cross--sections would decrease asymptotically like $1/\ln s$ , i.e. the relation
  \begin{equation}\label{bdindi1}
  \sigma_{diff}(s)/\sigma_{el}(s)\to 0
\end{equation}
will take place at $s \to \infty$.
  
It would be also interesting to speculate further and assume  the saturation of the bound Eq. (\ref{brfd1}). It would mean that an 
 asymptotic equipartition of the inelastic cross-section on diffractive and non-diffractive ones occurs.

\section*{Conclusion}    
 
 Thus, one can say that, at least, there is no  inconsistency between saturation of the unitarity limit leading to Eq. (\ref{rd}) and the  bound on the inelastic
diffractive cross--section in the case of reflective scattering, i.e. the reflective scattering limit  and the ratio  \[\sigma_{diff}(s)/\sigma_{inel}(s)\to const.\] at $s \to \infty$ can easily be reconciled. The energy-independent ratio $\sigma_{diff}(s)/\sigma_{inel}(s)$ is also consistent with the commonly accepted definition of the inelastic diffraction as a result of the Pomeron exchanges
and account for the recent experimental trends found at the LHC. 

Note, that this is not the case, if one assumes a mechanism resulting in the black disk limit at the asymptotic energies.  The black disk limit could usually be motivated by the eikonal models. Those models  reduce the range of the possible variation of a partial amplitude  by factor of 1/2. This is not in a direct inconsistency with the LHC data yet. However, the most recent analysis provides the strong indications on possibility  of the deviation  from the black disk limit\cite{alkin}. 
Thus, it seems difficult to conform the behavior of the inelastic diffraction  at the LHC energy range to the assumption on the black disk limit at $s\to\infty$.

The new LHC experiments
at higher  energies would be definitely helpful  for resolving the asymptotical dynamics of the inelastic diffraction and elastic scattering.

\end{document}